\documentstyle[preprint,aps]{revtex}
\begin{document}
%\input epsf
%@twocolumnfalse\endcsname
\title{Shot noise and coherent multiple charge transfer in 
superconducting quantum point-contacts}

\author{J.C. Cuevas, A. Mart\'{\i}n-Rodero and  A. Levy Yeyati}

\address{Departamento de F\'\i sica Te\'orica de la Materia Condensada C-V, 
Universidad Aut\'onoma de Madrid, E-28049 Madrid, Spain}

\date{\today}

\maketitle

\begin{abstract}

We analyze the shot noise in a voltage biased superconducting
quantum point-contact. Results are presented for the single channel case
with arbitrary transmission. In the limit of very low transmission it is 
found that the effective charge, defined from the noise-current ratio,
exhibits a step-like behavior as a function of voltage with
well defined plateaus at integer values of the electronic charge. This
multiple charge corresponds to the transmitted charge in a Multiple
Andreev Reflection (MAR) process. This effect gradually disappears for
increasing transmission due to interference between different MAR
processes.    

\end{abstract}

\pacs{PACS numbers: 72.70.+m, 74.50.+r, 74.80.Fp}     

%\vskip2pc]

\narrowtext

In the last few years much attention has been paid to the study of shot
noise in mesoscopic systems \cite{Jong1,Buttiker}. These time-dependent
current fluctuations are a consequence of the discreteness of the charge
carriers and their measurement can provide information on correlations
and charge of the individual carriers not available in usual
conductance experiments. In the case of a current
$I$ of uncorrelated carriers of charge $q$, the shot noise
reaches its maximum value $S = 2qI = S_{Poisson}$.  This result has been
recently used for detecting the fractional $e/3$ charge carriers by
measuring the ratio $S/2I$ in the fractional quantum Hall regime
\cite{fqhe}.

In low-transmitting normal-superconducting N-S structures a doubling of
the normal Poisson noise-current ratio has been predicted
\cite{Khlus,Jong2,Muzykantskii} due to Andreev processes where twice 
the electron charge is transmitted.
In S-N-S or S-I-S structures the situation is far more complex. In this
case, it is well established that the main processes contributing to
the current for subgap bias voltages are multiple Andreev reflections
(MAR) \cite{KBT}. At a given subgap voltage $V$ the current is mainly
carried by MAR processes of order $n \sim 2\Delta/eV$, in which 
a net charge of $ne$ is transferred. One would then expect an increase of the
noise-current ratio roughly as $1/V$ for decreasing bias. This
qualitative behavior has been recently confirmed experimentally by
Dieleman et al. \cite{Dieleman} for a S-I-S tunnel junction.
These authors give an explanation of the observed shot noise
enhancement due to MAR within the framework of the semi-classical
theory of Ref. \cite{KBT}. 

In large tunnel junctions a quantitative comparison between theory and 
experiment is prevented by unavoidable uncertainties in the junction 
geometrical structure. On the other hand, there has been in recent years
a large progress in the fabrication of superconducting point contacts with
a reduced number of conducting channels with rather controllable
transmission \cite{Scheer1,Scheer2}. The subgap structure (SGS) in the I-V
characteristics of these systems has been quantitatively described with
high accuracy using fully quantum mechanical theories of transport through
a single channel \cite{Averin1,tocho}. One would expect that a
similar quantitative agreement between theory and experiment could be
reached also in the case of noise. 

So far the analysis of current fluctuations in a superconducting single 
channel contact has been restricted to a few special cases. Thus, the
excess noise ($eV \gg \Delta$) for a perfect transparent channel has
been obtained in Ref. \cite{Hessling}. The noise has also been analyzed
for zero voltage and arbitrary transmission in  Ref. \cite{us}, while
the perfect transmission and finite voltage case has been addressed in
Ref. \cite{Averin2}.  

The aim of this work is to analyze the shot noise for the whole range of
transmissions and voltages within the same microscopic model used for
the calculation of the current in Ref. \cite{tocho}. This analysis would
allow to answer the question on whether a well defined $q=S/2I$ transmitted
effective charge can be associated to the processes giving rise to the SGS 
in the current.
We shall explicitly show that this is only possible in the low 
transmission regime where the effective charge tends to \emph{integer
multiples} of the electron charge $n=1 + Int[2\Delta/eV]$.

We shall consider the case of a superconducting quantum point-contact
(SQPC), i.e. a short ($L \ll \xi_0$) mesoscopic constriction between two
superconducting electrodes with a constant applied bias voltage $V$. 
For the range $eV \sim
\Delta$ one can neglect the energy dependence of the transmission
coefficients and all transport properties can be expressed as a
superposition of independent channel contributions \cite{Averin3}. 
Thus, we will concentrate in analyzing a single channel model which can be
described by the following Hamiltonian \cite{tocho}

\begin{equation}
\hat{H}(\tau) = \hat{H}_{L} + \hat{H}_{R} +
\sum_{\sigma} \left( t e^{i \phi(\tau)/2} c^{\dagger}_{L \sigma} c_{R \sigma} 
+ t^{*} e^{-i \phi(\tau)/2} c^{\dagger}_{R \sigma} c_{L \sigma} \right),
\end{equation}

\noindent
where $\hat{H}_{L,R}$ are the BCS Hamiltonians for the left and
right uncoupled electrodes, $\phi(\tau) =  \phi_{0} + 2eV\tau/
\hbar$ is the time-dependent superconducting phase difference, which
enters as a phase factor in the hopping terms describing electron
transfer between both electrodes. The 
normal transmission coefficient, $\alpha$, can be varied between 
$0$ and $1$ as a function of the hopping parameter 
$t$ (see ref. \cite{tocho} for details).
Within this model the current operator is given by

\begin{equation}
\hat{I}(\tau)= \frac{i e}{\hbar}  \sum_{\sigma} \left( t e^{i
\phi(\tau)/2} c^{\dagger}_{L\sigma}(\tau) c_{R \sigma}(\tau) - \;
t^{*} e^{-i \phi(\tau)/2} c^{\dagger}_{R \sigma}(\tau) 
c_{L \sigma}(\tau) \right) .
\end{equation}

The current-noise spectral density is defined as

\begin{equation}
S(\omega,\tau) = \hbar \int d \tau^{\prime} \; e^{iw \tau^{\prime}}  
<\delta \hat{I}(\tau + \tau^{\prime}) \delta \hat{I}(\tau)
+ \delta \hat{I}(\tau) \delta\hat{I}(\tau + \tau^{\prime})>
\equiv \hbar \int d \tau^{\prime} \; e^{iw\tau^{\prime}} \; 
K(\tau, \tau^{\prime}) ,
\end{equation}

\noindent
where $\delta \hat{I}(\tau)= \hat{I}(\tau) - <\hat{I}(\tau)>$ is the
time-dependent fluctuations in the current.

In the case of a voltage biased superconducting contact both 
$<\hat{I}(\tau)>$ and
$S(\omega,\tau)$ contains all the harmonics of the Josephson frequency
$\omega_0 = 2eV/\hbar$, i.e. $<\hat{I}(\tau)> = \sum_n I_n 
\exp[in\omega_0 \tau]$. 
These quantities can be expressed in terms of non-equilibrium Green
functions in a superconducting broken symmetry or Nambu representation
$\hat{G}^{+-}_{ij}(t,t^{\prime})$ and $\hat{G}^{-+}_{ij}(t,t^{\prime})$
where $i,j \equiv L,R$ defined as

\begin{equation}
\hat{G}^{+-}_{i,j}(\tau,\tau^{\prime})= i \left(
\begin{array}{cc}
<c^{\dagger}_{j \uparrow}
(\tau^{\prime}) c_{i \uparrow}(\tau)>   &
<c_{j \downarrow}(\tau^{\prime}) c_{i \uparrow}(\tau)>  \\
<c^{\dagger}_{j \uparrow}(\tau^{\prime})
c^{\dagger}_{i \downarrow}(\tau)>  &
<c_{j \downarrow}(\tau^{\prime})
 c^{\dagger}_{i \downarrow}(\tau)>
\end{array}  \right)  ,
\end{equation}

\noindent
and obey the relation $\hat{G}^{-+}_{i,j}(\tau,\tau^{\prime})
= \left[\hat{G}^{+-}_{j,i}(\tau,\tau^{\prime})\right]^{\dagger}$.

Then, the mean current and the kernel $K(\tau,\tau^{\prime})$ in the noise
spectral density are given by

\begin{eqnarray}
<\hat{I}(\tau)> & = & \frac{e}{\hbar} \; Tr \left[ \hat{\sigma}_z 
\left( \hat{t}(\tau) \hat{G}^{+-}_{RL}(\tau,\tau) -
\hat{t}^{\dagger}(\tau)
\hat{G}^{+-}_{LR}(\tau,\tau) \right) \right] \nonumber \\
K(\tau,\tau^{\prime}) & = & \frac{e^2}{\hbar^2} \left\{ Tr \left[
\hat{t}^{\dagger}(\tau) \hat{G}^{+,-}_{LL}(\tau, \tau^{\prime}) 
\hat{t}(\tau^{\prime}) \hat{G}^{-,+}_{RR}(\tau^{\prime},\tau) +
\hat{t}(\tau) \hat{G}^{+,-}_{RR}(\tau, \tau^{\prime})
\hat{t}^{\dagger}(\tau^{\prime}) \hat{G}^{-,+}_{LL}(\tau^{\prime},\tau) -
\right. \right. \nonumber \\
& & \left. \left. \hat{t}^{\dagger}(\tau) 
\hat{G}^{+,-}_{LR}(\tau, \tau^{\prime})
\hat{t}^{\dagger}(\tau^{\prime}) \hat{G}^{-,+}_{LR}(\tau^{\prime},\tau) -
\hat{t}(\tau) \hat{G}^{+,-}_{RL}(\tau, \tau^{\prime})
\hat{t}(\tau^{\prime}) \hat{G}^{-,+}_{RL}(\tau^{\prime},\tau) \right] 
+ (\tau \rightarrow \tau^{\prime}) \right\} ,
\end{eqnarray}

\noindent
where $\hat{\sigma}_z$ is the Pauli matrix, $Tr$ denotes the trace in the
Nambu space and $\hat{t}$ is the hopping in this representation

\begin{equation}
\hat{t} = \left(
\begin{array}{cc}
 t e^{i \phi(\tau)/2}  &   0      \\
  0                    &   -t^* e^{-i \phi(\tau)/2}
\end{array} \right)  .
\end{equation}

In the expression (5) for the kernel we have factorized the two body correlation
functions following a mean field BCS decoupling scheme \cite{us}. 
The problem of evaluating the Fourier components
of $I$ and $S$ can be reduced to the calculation of the Fourier
components of the Keldysh Green functions. An efficient algorithm for
this evaluation can be found in Ref. \cite{tocho}. We shall restrict our
attention to the zero-frequency dc component of the noise $S \equiv
\overline{S(0,\tau)}$ at zero temperature.

Fig. 1 illustrates the behavior of $S$ as a function of $V$ for
different values of the normal transmission $\alpha$. For comparison,
the dc component of the current is also shown. As can be observed, the
more noticeable features in the shot noise are: (i) the presence of a
strongly pronounced subgap ($V \le 2\Delta$) structure, which persists
up to transmissions close to one (in the dc current this structure is
only pronounced for low transmissions). (ii) In the low 
transparency limit the shot noise subgap structure consists of a series of
steps at voltages $eV_n=2\Delta/n$ ($n$ integer) as in the case of the
dc current. (iii) For higher transmissions there is a steep increase in
the noise at low voltages. (iv) For
perfect transmission the shot noise is greatly reduced. (v) In the large
voltage limit there is an excess noise with respect to the normal case.

Let us start by analyzing the low transmission regime. In this case it
turns out that the electronic transport is well described by a 
sequential tunneling
picture \cite{dot} and the current can be written as the sum of
tunneling rates for each multiple Andreev process times the transmitted
charge, i.e. $I_0(V) = e \sum n \Gamma_n(V)$ with
$\Gamma_n = (2/h) \int d\omega \; R_n(\omega)$. The probability of an
$n$th-order Andreev process $R_n$ is given by \cite{tocho}

\begin{equation}
R_n(\omega) = \frac{\pi^2 \alpha^n}{4^{n-1}}
\left[ \prod^{n-1}_{i=1} | p(\omega - ieV) |^2 \right]
\rho(\omega - neV) \rho(\omega) \; \; ; \; \omega \in [\Delta,neV-\Delta],
\end{equation}

\noindent
where $\rho(w) = |\omega|/\sqrt{\omega^2-\Delta^2}$ is the dimensionless
BCS density of
states and $p(\omega) \sim \Delta/\sqrt{\Delta^2-\omega^2}$ is the
Cooper pair creation amplitude. Expression (7) for $R_n$ clearly
displays the different ingredients in a MAR processes, i.e. it is
proportional to the initial and final density of states, to the
probabitlity of creating $n-1$ Cooper pairs and to the probability of a
quasiparticle crossing $n$ times the interface ($\alpha^n$).
In this regime the shot noise adopts an appealing form in terms of $R_n$

\begin{equation}
S = \frac{4e^2}{h} \int d\omega \; \left\{ \sum^{\infty}_{n=1} n^2 R_n -
\left( \sum^{\infty}_{n=1} n R_n \right)^2 \right\} .
\end{equation}

\noindent
This expression corresponds to the fluctuations of a random variable
(the current) having a multinomial distribution. This is a
generalization of the simple binomial distribution one finds in a N-N
contact, which leads to the well known $\alpha(1-\alpha)$
behavior \cite{Jong1,Buttiker,Khlus}.

This analysis shows that in the limit of vanishing transmission 
only the lowest order process with a non-zero probability contributes to
the current at a given bias voltage. As a   
consequence the effective charge shows a step-like behavior described
by the formula $q/e = S/2eI = 1 + Int \left[2\Delta/eV \right]$.   
The inset in figure 2(b) shows the
effective charge for $\alpha=0.01$ together with the
step-like function corresponding to the $\alpha \rightarrow 0$ limit. 
It is also shown in this figure
the comparison between the tunnel approximation given by Eq. (8)
and the exact result for a transmission $\alpha=0.1$. Notice that even for
this small transmission value there are deviations from the simple
step-like behavior which are increasingly pronounced when the voltage is
reduced. These deviations are produced by the contribution of more than one
MAR process at a given voltage. 

As the transmission increases the sequential tunneling
picture breaks down due to the interference between different MAR processes
contributing to the current at any voltage. The
charge quantization found in the tunnel regime progressively disappears and
is eventually washed out when approaching perfect transmission. This is
illustrated in Fig. 2(a).

For intermediate transmissions the analysis of the shot noise becomes rather
involved not only due to the interference between different
processes but also to the increasing contribution of 
fluctuations of higher harmonics of the current. On the other hand,
in the perfect transmission limit, the analysis is again simplified due to the
absence of backscattering. Within our theory one obtains
a simple expression for the shot noise in this limit

\begin{equation}
S = \frac{8e^2}{h} \int d\omega \; \left[ \sum^{\infty}_{n=0}
R_n (1-R_n) \right] \left[1+2 \sum^{\infty}_{k=1} \prod^k_{l=1}
|r(\omega+leV)|^2 \right] ,
\end{equation}

\noindent
where $R_n$ are the multiple Andreev reflection probabilities given by
$R_n(\omega)= \prod^n_{m=0} |r(\omega-meV)|^2  ; \; \omega
\in [neV,(n+1)eV]$ and $r(\omega)=(\omega+i\sqrt{\Delta^2-\omega^2}) 
/ \Delta$ is the Andreev reflection amplitude at an N-S interface. 
This expression can be shown to be
equivalent to the result of Averin and Iman in Ref. \cite{Averin2}.
The great reduction of noise that can be observed
in Fig. 1 for perfect transmission is a consequence of having an 
Andreev reflection probability equal to one inside the gap.

It is also interesting to analyze the large voltage limit, in which the 
zero-frequency noise behaves as
$\lim_{V \rightarrow \infty} S = (4e^2/h) \alpha(1-\alpha) V + S_{exc}$ 
, i.e. the shot-noise of a normal contact with transmission $\alpha$ plus
and ``excess noise'' $S_{exc}$. The excess noise as a function of 
transmission is shown in Fig. 3. $S_{exc}$ has the same physical
origin as the excess current ($I_{exc}$), which arises from the
contribution of the first order Andreev process. One should remark that
the approach to the asymptotic value is much slower for the shot noise
than for the current. We obtain
that at zero temperature $S_{exc}$ is twice the excess noise of a N-S
contact \cite{Khlus} with the same transmission. 
In particular, this relation yields $S_{exc} = 2/5 e I_{exc}$ for the
perfect ballistic case in agreement with Ref. \cite{Hessling}.

A word of caution should be said about the validity of this theory in
the limit of extremely small bias voltage. 
In this limit there is another energy scale determined by the
inelastic relaxation rate $\eta$ (which is a small fraction of $\Delta$)
playing a role in the theory (see Ref. \cite{us}). This finite relaxation 
rate introduces a cut-off in the MAR processes which determines the
behavior of current \cite{tocho} and noise when $eV \ll \eta$. 
The precise behavior at $V \rightarrow 0$ thus depends on the actual
value of $\eta$.

In conclusion, we have analyzed theoretically the shot noise in a single
channel SQPC for arbitrary transmission and bias voltage.
We have shown that 
the shot noise can be much larger than the Poisson noise ($S_{Poisson}=2eI$)
due to the occurrence of multiple Andreev reflections in which multiple
charge quanta are transferred. In the tunnel
regime, the effective charge $q(V)=S/2I$ shows a step-like behavior
which is a signature of the coherent transmission of multiple electronic 
charges. Our results are consistent with the  
available experimental data of Dieleman \emph{et al.} \cite{Dieleman}.
These authors used large tunnel junctions for which a mean transmission
of $\alpha=0.17$ was estimated. Although a direct comparison with theory
is difficult due to uncertainties in the junction structure,
the experimental results for shot noise and the
effective charge are in qualitative agreement with the results of our
Figs. 1-2 for this transmission range. On the other hand, the predictions
presented in this letter are amenable to direct experimental test using
state of the art techniques for fabricating atomic-size contacts
\cite{Scheer1,Scheer2}. In view of the remarkable   
agreement between theory and experiments for the current-voltage
characteristics found in these systems,
one would expect a similar quantitative agreement 
when measuring current-current fluctuations.

\acknowledgements
The authors would like to thank C. Urbina and E. Scheer for useful
discussions. This work has been supported by the Spanish CICYT under
contract No. PB97-0044.

\begin{figure}
\caption{(a) Current-voltage and (b) noise-voltage characteristics 
for different transmissions at zero temperature. The values of the
transmission are the same in both panels.}
\end{figure}

\begin{figure}
\caption{(a) Effective charge as a function of voltage for different
transmissions. (b) Comparison between the exact result (full line) 
and the tunnel approximation given by Eq. (8) (open
circles) for $\alpha=0.1$. The inset shows 
the exact result (full line) for $\alpha=0.01$ and the step-like function 
(dashed line) corresponding to the $\alpha \rightarrow 0$ limit.}
\end{figure}

\begin{figure} 
\caption{Excess noise as a function of transmission at zero
temperature.}
\end{figure} 

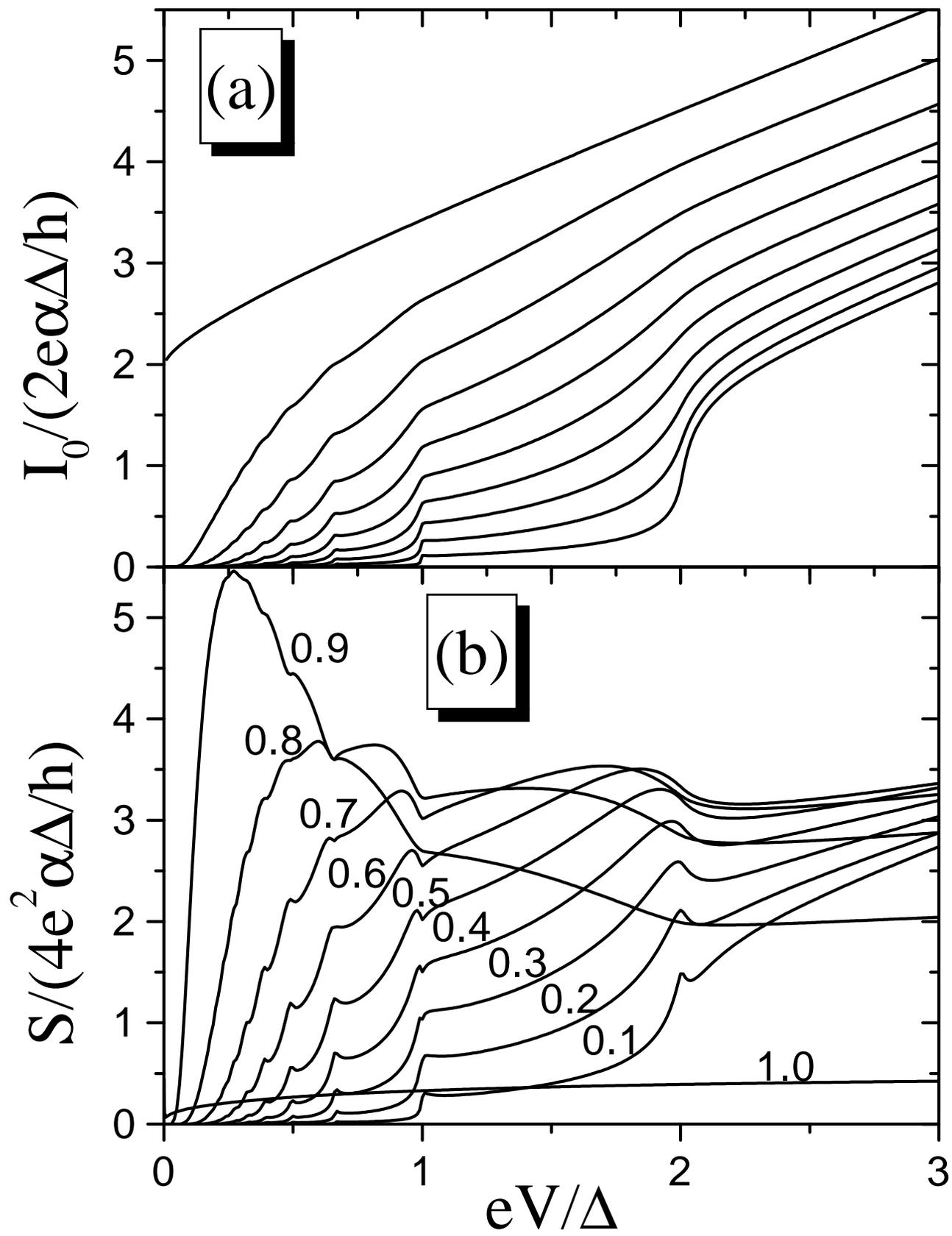
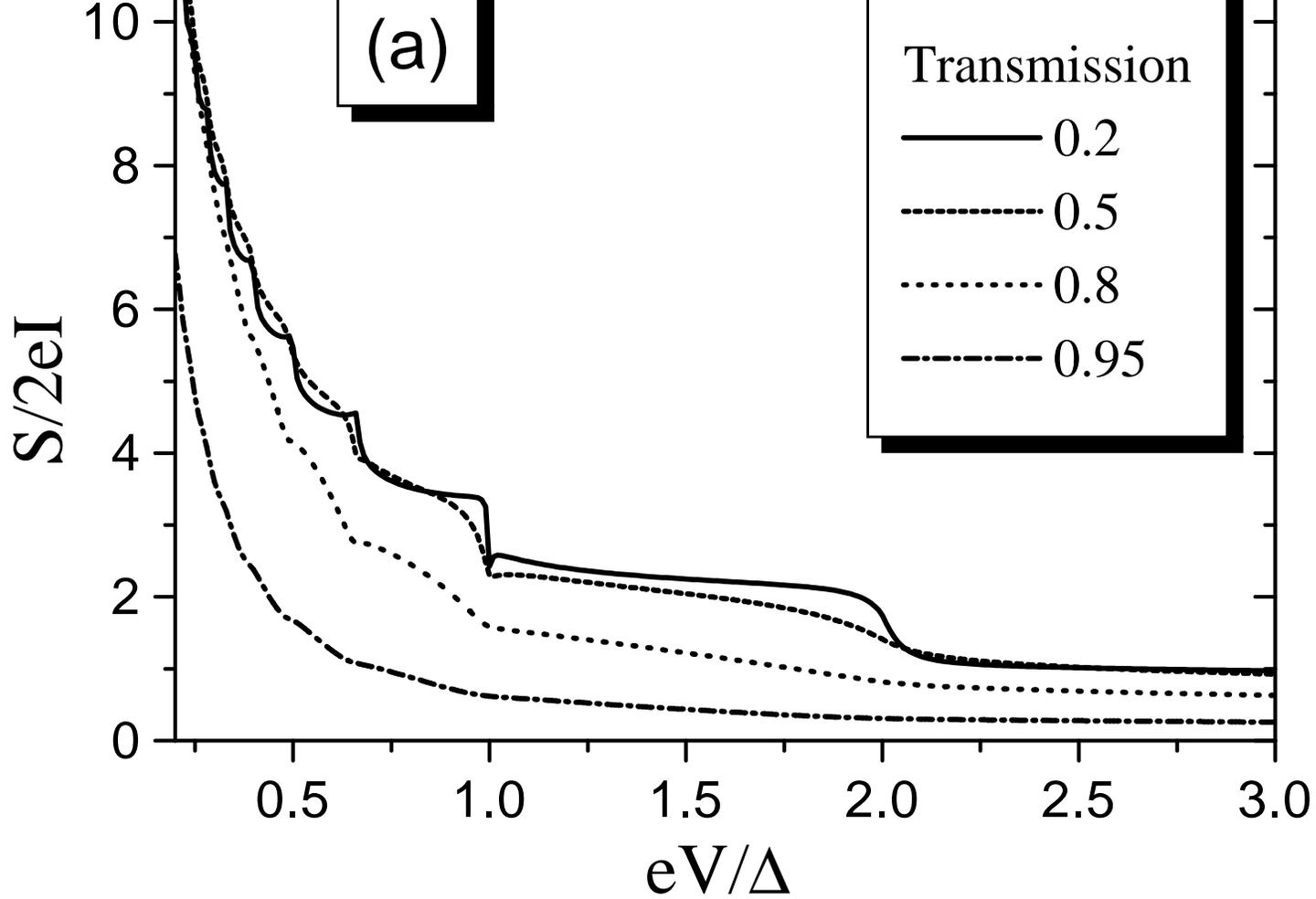
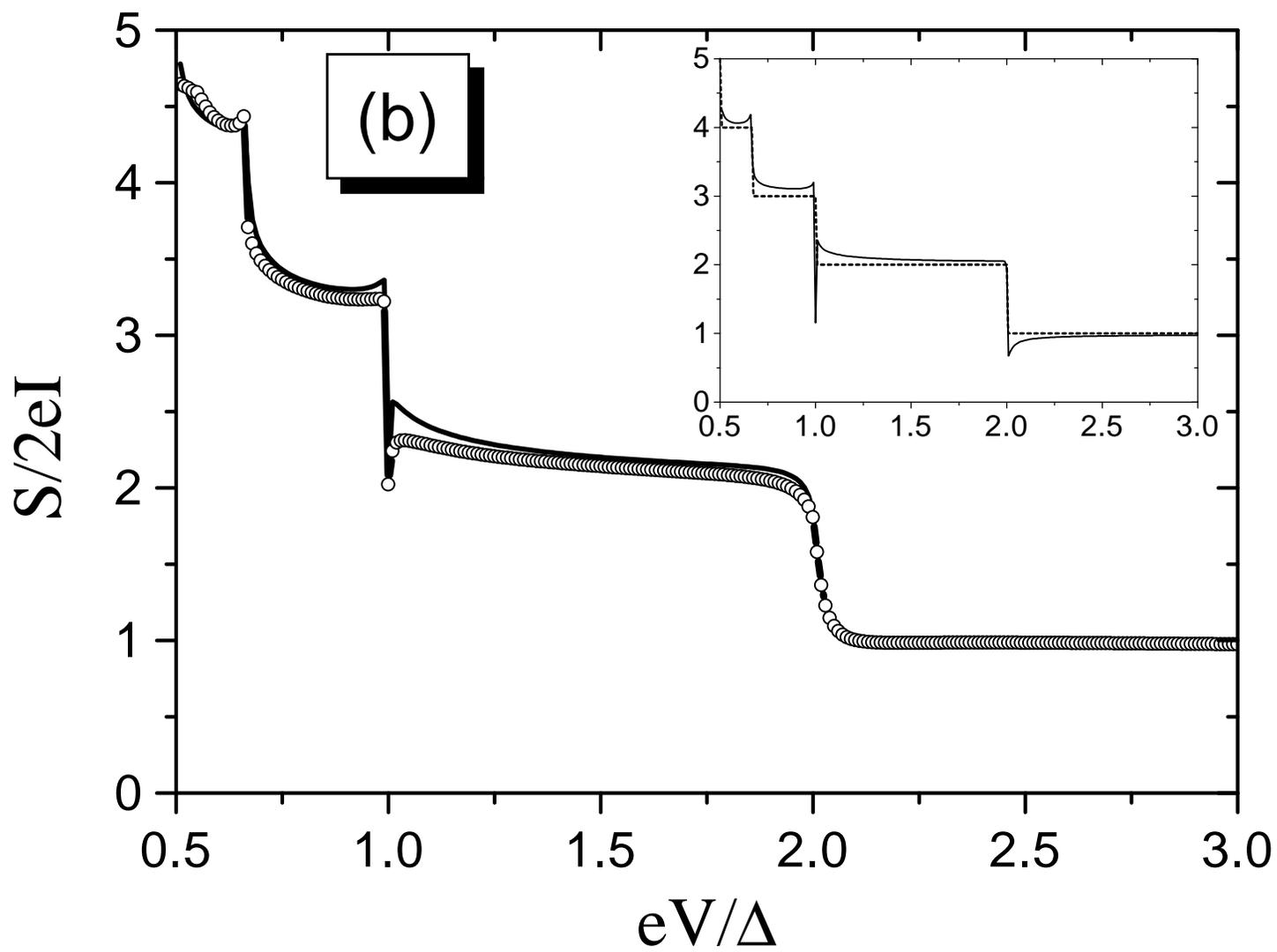
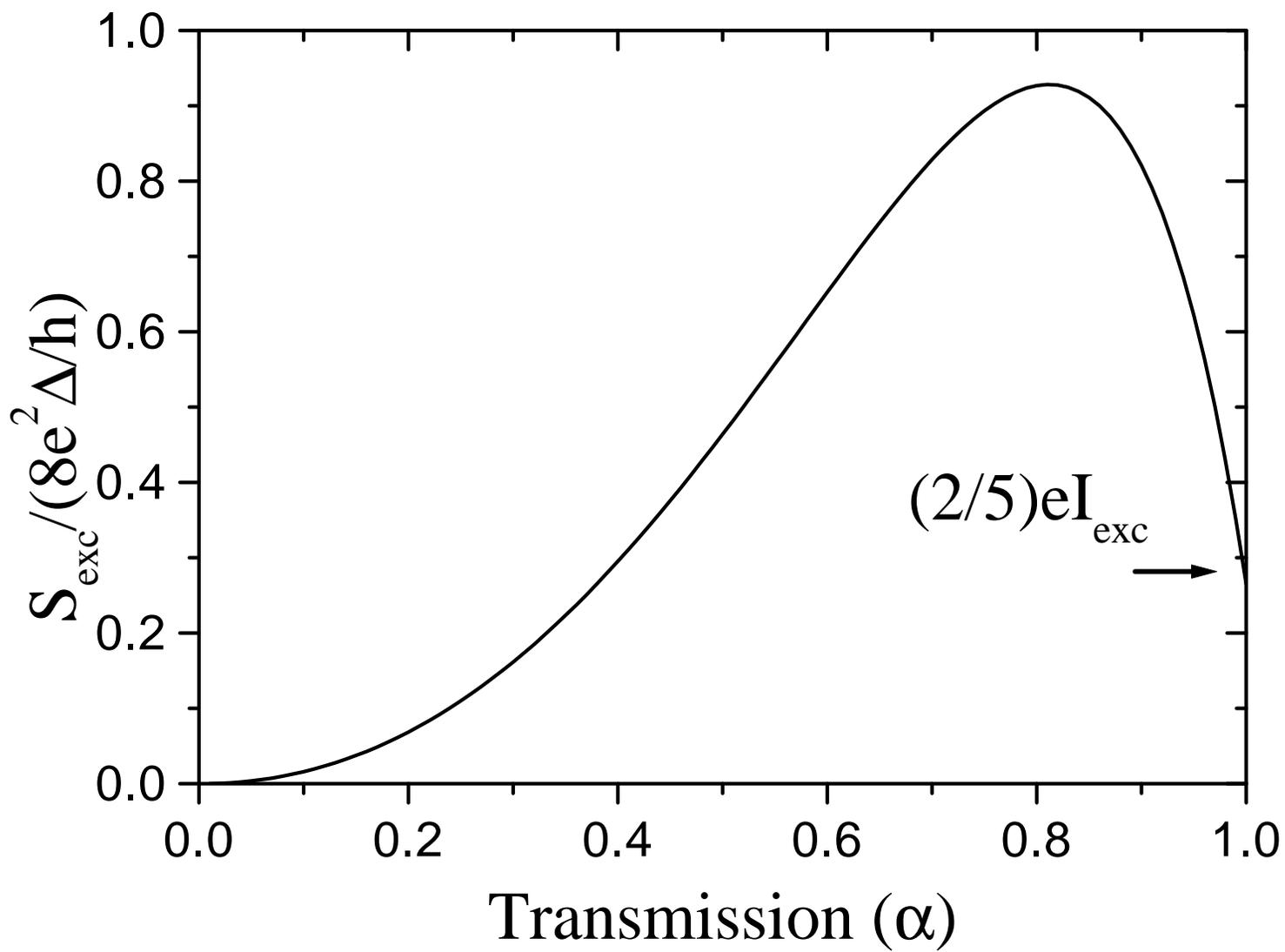
\end{document}